\documentclass[10pt]{article}
\usepackage[OE]{express}
\usepackage{cite}
\usepackage[colorlinks=true,urlcolor=blue,linkcolor=black,citecolor=black,breaklinks=true,bookmarks=false]{hyperref}

\begin{document}

\title{Photothermal effects in ultra-precisely stabilized tunable microcavities}

\author{Johannes F.\ S.\ Brachmann,$^{1,2}$ Hanno Kaupp,$^{1,2}$ Theodor W.\ H\"ansch,$^{1,2}$ and David Hunger$^{1,2,*}$}

\address{$^1$Fakult{\"a}t f{\"u}r Physik, Ludwig-Maximilians-Universit{\"a}t, Schellingstra{\ss}e~4, 80799~M{\"u}nchen, Germany\\
$^2$Max-Planck-Institut f{\"u}r Quantenoptik,  Hans-Kopfermann-Str.~1, 85748~Garching, Germany}

\email{$^*$david.hunger@physik.lmu.de}


\begin{abstract}
We study the mechanical stability of a tunable high-finesse microcavity under ambient conditions and investigate light-induced effects that can both suppress and excite mechanical fluctuations. As an enabling step, we demonstrate the ultra-precise electronic stabilization of a microcavity. We then show that photothermal mirror expansion can provide high-bandwidth feedback and improve cavity stability by almost two orders of magnitude. At high intracavity power, we observe self-oscillations of mechanical resonances of the cavity. We explain the observations by a dynamic photothermal instability, leading to parametric driving of mechanical motion. For an optimized combination of electronic and photothermal stabilization, we achieve a feedback bandwidth of $500\,$kHz and a noise level of $1.1 \times 10^{-13}\,$m rms.
\end{abstract}




\section{Introduction}
Tunable Fabry-Perot microcavities offer enhanced light-matter interaction in combination with open access and full control of the mirror separation \cite{Trupke05,Hunger10b,Toninelli10,Muller09,Dolan10,Barbour11}. The design facilitates the introduction of samples into the cavity and enables optimal spectral and spatial overlap between the sample and the optical mode. This is beneficial for a broad range of applications, e.g.\ in the fields of cavity quantum electrodynamics \cite{Colombe07,Muller09,Toninelli10,Barbour11,Steiner13,Albrecht13,Kaupp13,MiguelSanchez13,Johnson15}, cavity optomechanics \cite{Favero09,Stapfner13,FlowersJacobs12}, and cavity-enhanced microscopy \cite{Mader15,Huemmer16}. In particular, the approach to scan one of the mirrors to record spatially resolved measurements \cite{Toninelli10,Greuter14,Mader15,Huemmer16} opens up novel possibilities e.g.\ for studies of heterogeneous samples. However, a scannable cavity requires sophisticated control to maintain resonance conditions with a probe laser or optical transitions of a sample. If operation under ambient conditions in air is envisaged, the cavity setup will be exposed to acoustic and structure-borne noise, as well as thermal drifts and pressure variations. This leads to fluctuations of the relative mirror position and the optical path length, which define the resonance frequency of the cavity.

For a scannable fiber-based microcavity under ambient conditions on an optical table, we typically observe a mechanical noise spectrum with significant spectral components extending up to a few kHz, and an average noise level of $0.3 - 1$\,nm rms within 1\,s. Compared to e.g.\ monolithic Fabry-Perot fiber cavities in vacuum \cite{Colombe07,Gallego16} or mechanically and acoustically isolated scannable cavities \cite{Greuter14}, the fluctuations are $2-3$ orders of magnitude larger. On timescales of minutes and hours, thermal and mechanical drifts lead to position noise with peak-to-peak amplitudes of several hundred nanometers. This is in contrast to the narrow resonance condition of cavities with high finesse $F$. When probing e.g. with a laser with fixed wavelength $\lambda$, a change of the cavity length by $\delta z = \lambda/(2 F)$ corresponds to crossing the full width half maximum (FWHM) of the cavity resonance, and $\delta z$ is thus the resonance width in the spatial domain. For a finesse $F\sim 10^5$, this amounts to a length change of $\delta z\sim 1\,$pm only.

In this work we report the stabilization of a high-finesse fiber-based microcavity \cite{Hunger10b} with a dedicated electronic lock as well as with photothermal feedback to a level of $114\,$fm rms. Due to the high cavity finesse of 172000, we can resolve mechanical fluctuations with amplitude spectral densities down to $2\times 10^{-17}$m/$\sqrt{\textrm{Hz}}$ for frequencies above 1\,MHz.
We study photothermal self-stabilization as an efficient way to achieve high-bandwidth feedback, significantly extending the frequency range available for mechanical feedback, which is in general limited by mechanical resonances of the mounting structure.
Furthermore, we find that already at moderate intracavity power, parametric excitation of mechanical modes of the cavity mirrors occurs, which limits the useful range of intracavity powers.
We discuss the influence of photothermal and radiation pressure effects on the cavity stability and demonstrate a significant improvement of an optimized electronic lock by photothermal self-stabilization.\\


\section{Experimental setup}

\begin{figure*}
\centering\includegraphics[width=.9\textwidth]{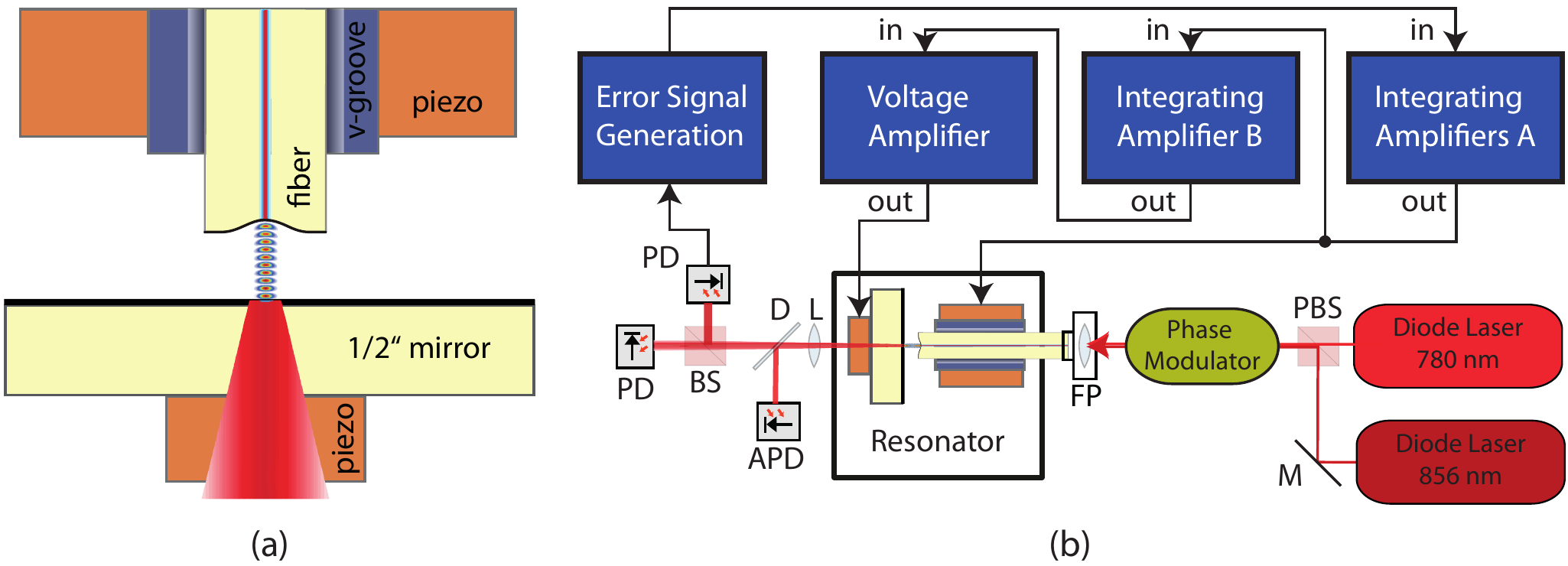}
\caption{(a) Schematic of the microresonator consisting of a micro-machined and mirror-coated fiber mounted on a piezo shear plate and a 1/2" planar mirror with a stacked piezo.
(b) Scheme of the experimental setup. For Pound-Drever-Hall stabilization, the light at 856\,nm is phase modulated using a fiber-coupled electro-optical modulator.
Modulation sidebands were also used to determine the resonator linewidth. PBS: Polarizing beam splitter, M: mirror, FP: Fiber port, PD:  Photo diode, BS: nonpolarizing beam splitter, APD: Avalanche photo diode, D: dichroic mirror, L: Lens.}
\label{fig:cavity_setup}
\end{figure*}

Figure \ref{fig:cavity_setup}(a) shows a schematic of the optical cavity. A plane-concave resonator geometry is realized between a laser-machined and mirror-coated optical fiber end-facet \cite{Hunger12} and a 1/2" plane mirror. 
The plane mirror can be angle adjusted with a Gimbal mount and coarse positioned with a manual micropositioning stage.
A stacked piezo-electric transducer moving the 1/2" mirror over up to 1.5 free spectral ranges is used to tune the cavity length to resonance, while a high-bandwidth monolayer piezo shear plate on the fiber side allows to tune the length by about 5\,nm -- which is slightly larger than the amplitude of the fast mechanical fluctuations occuring in the setup. A low expansion coefficient of 0.5\,nm/V of the shear plate relaxes the noise requirements for the feedback electronics.
We observe the lowest mechanical resonance of the mounting structure for the cavity fiber at 25\,kHz, which determines the maximal bandwidth of the electronic feedback loop. In more optimized designs, feedback bandwidths of up to 200\,kHz have been achieved \cite{Briles10}. To demonstrate the robustness of the subsequent stabilization, the cavity is operated on a non-floating optical table in air without any acoustic shield.

Figure \ref{fig:cavity_setup}(b) shows a schematic of the electronic and optical setup. Two grating-stabilized diode lasers at 780\,nm and 856\,nm are used to probe and stabilize the cavity, respectively. In cavity transmission, the two wavelengths are separated by a dichroic mirror and detected with specific photodiodes. We determine the finesse from a measurement of the cavity's quality factor and length, where the latter is determined from the mirror-distance-dependent cavity transmission of both lasers. By use of modulation sidebands, we can measure the linewidth in units of frequency.

The dielectric coating of the microcavity mirrors is optimized for 780\,nm. At this wavelength, we observe a maximum resonator finesse of $F=172000 \pm 10000$ and a linewidth of 115\,MHz for a resonator length of 7.5\,$\mu$m, at which the experimental data was acquired. The fundamental cavity resonance is split into a polarization doublet with $\sim 1\,$linewidth separation due to geometric birefringence \cite{Uphoff15}. In the experiment, we set the laser polarization such that we probe only one of the resonances.

The resonator length is actively stabilized to the laser at 856\,nm, which serves as a frequency reference and thus as a length reference. The laser can serve as a reasonable reference also without stabilization, since the length of the external cavity of the laser is $\sim 10^3$ times larger than that of the microcavity, such that the same amount of mechanical fluctuations lead to a $\sim 10^3 -$times smaller frequency noise of the laser.
We frequency modulate the light with an EOM at 3.6\,GHz to generate an error signal and use a Pound-Drever-Hall (PDH) stabilization scheme \cite{Drever83,Black01} in cavity transmission. This reduces error signal offset drifts caused by a fluctuating parasitic interferometer which is formed within the cavity fiber between the micro mirror and the angle-cleaved incoupling facet. At 856\,nm, the finesse is 3000, such that the cavity linewidth is larger than the modulation frequency, and the carrier is transmitted with the sidebands. The reduced finesse enables capturing the fluctuating cavity resonance, which was not possible for 780\,nm due to the finite feedback amplitude at maximum bandwidth. The light at 856\,nm is detected by two photodiodes, one with $> 4\,$GHz bandwidth and 72\,dB amplification for error signal generation, the other for monitoring. The error signal is fed to an amplifier consiting of two added integrators (amplifiers A in Fig.\ 1(b)) with different cut-off frequencies in series with a variably tunable passive lowpass filter. This allows to optimize the feedback loop transfer function. In order to compensate drifts which exceed the maximum amplitude of the fiber-side piezo shear plate, a second integrating amplifier (amplifier B) of very low bandwidth (15\,Hz) is connected to a voltage amplifier to drive the stacked piezo. The connection of the integrator's input directly to the output of the wide-bandwidth loop reliably compensates all drifts.\\

\begin{figure}
\centering\includegraphics[width=.8\textwidth]{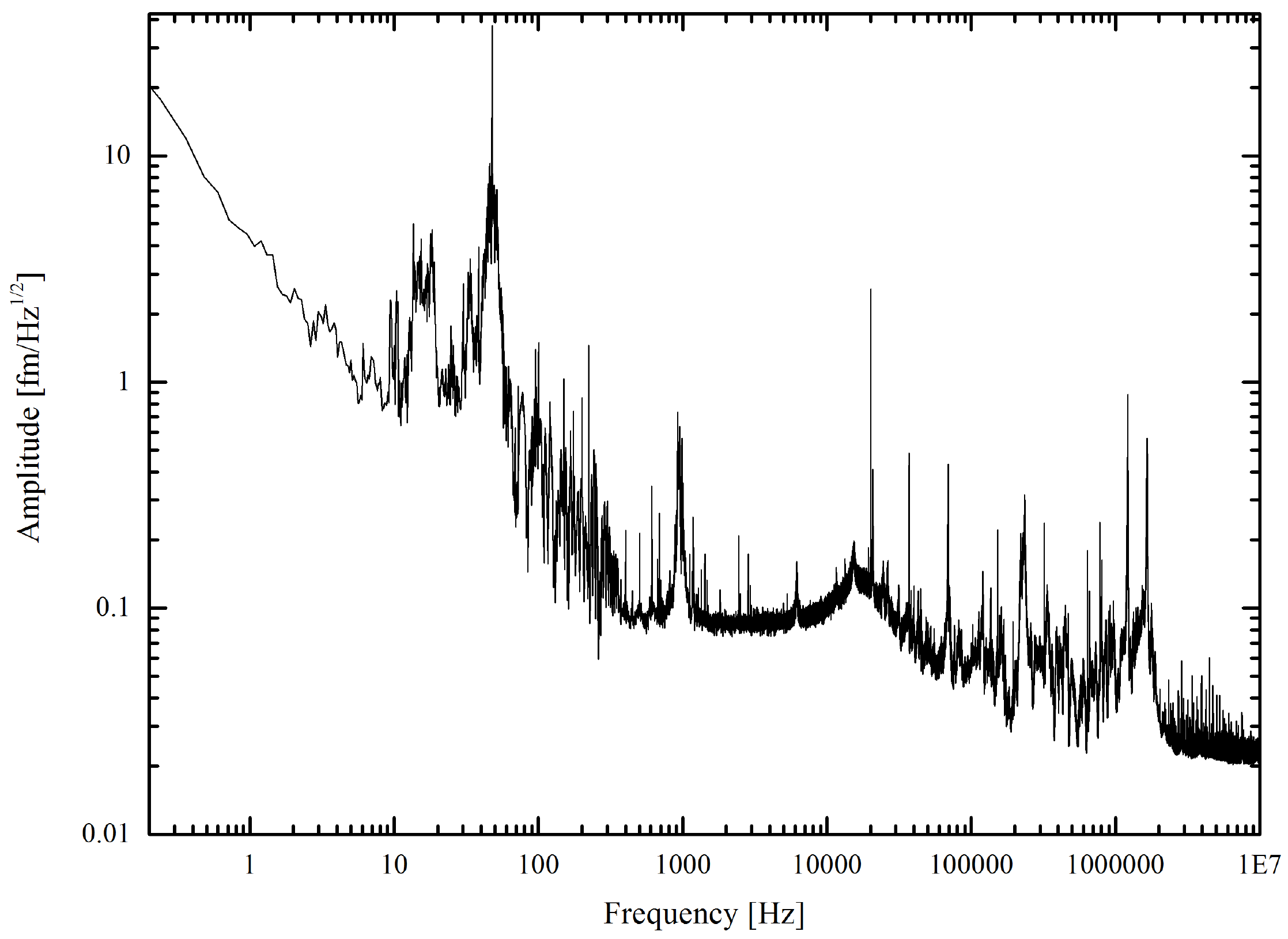}
\caption{Noise spectral density of the relative mirror separation for the actively stabilized resonator. The rms precision of length stability is 153\,fm (rms), evaluated in the frequency interval from 0.1\,Hz to 10\,MHz. We conservatively estimate an error of \,+100, -50\,fm for this value. In the optimized setting, the feedback bandwidth was set to 1.1\,kHz.
}
\label{fig:electronic_noise_spectrum}
\end{figure}


\section{Performance of the electronic stabilization}

To characterize the performance of our electronic feedback loop, we tune the wavelength of the laser at 780\,nm to the slope of a resonance of the stabilized cavity. In this manner, fluctuations of the mirror separation $z(t)$ are translated into amplitude fluctuations of the transmitted laser power $P(t)$, which we measure with a low-noise APD with 50\,MHz bandwidth and a 12\,bit oscilloscope. A Fourier transform of the recorded time traces gives the amplitude spectral density in units of V$/\sqrt{\mathrm{Hz}}$. To convert these spectra to length units, we consider the cavity resonance as a function of mirror separation $P(z)=P_r (\delta z/2)^2[z(t)^2+(\delta z/2)^2]^{-1}$, where $\delta z = \lambda/(2F)=2.27\,$pm is the measured FWHM in the spatial domain, and $P_r$ is the transmitted power on resonance. We measure $P(t)$ and want to infer $z(t)$ from it, so we invert $P(z)$ on one half of the resonance and obtain $z(P)=(\delta z/2) \sqrt{P_r/P(t)-1}$. Since we evaluate FFT spectra rather than time-domain data, we cannot directly apply this formula. Instead, we linearize $z(P)$ around the point of measurement, which is set close to $P=P_r/2$. We use the slope $\alpha = d z(P)/dP$ to calculate the amplitude spectrum $a(f)$ of the mechanical noise from the amplitude spectrum $P(f)$ of the transmitted laser power, $a(f)=\alpha P(f)$. During the time needed for multiple averaging, the probe and lock laser drift relative to each other, and the point of measurement varies. Therefore we use an averaged value $\bar{\alpha}=\int_{P_1}^{P_2}\alpha(P)dP$ for the calibration, with $P_1 = 0.4 P_r$ and $P_2 = 0.8 P_r$ representing the typical variation observed during measurements. The estimated uncertainties of evaluated cavity stabilities are dominated by this calibration uncertainty.

Figure \ref{fig:electronic_noise_spectrum} shows the amplitude spectral density of the fluctuations of the relative mirror separation under optimal electronic stabilization conditions. The intracavity power at 780\,nm is kept at 50\,mW for this measurement to maintain a Lorentzian resonance shape (see below). The main features of the spectrum are $1/f$ noise below 10 Hz, which we attribute to error signal offset fluctuations as well as frequency and amplitude noise of the free-running lasers; furthermore, a servo bump from the slow feedback on the large mirror at 47 Hz, and some sharp resonances at higher frequency originating from thermally excited mechanical resonances and electronic noise.

We note that the frequency and amplitude noise of the lock and probe lasers lead to a limitation of the achievable cavity stability. To estimate this limit, we have measured the power stability of both lasers and analyzed the corresponding noise spectra. They do not show marked spectral features and are smoothly decaying with frequency, and the relative rms noise was $< 5 \times 10^{-4}$ for both lasers. Compared to the relative amplitude noise observed for best cavity stability conditions, which amounts to $5 \times 10^{-2}$ rms at the maximal slope of the microcavity resonance, the laser amplitude noise remains two orders of magnitude smaller and should thus not contribute notably to the noise spectra.
We have also estimated the frequency noise of both lasers by coupling them to a passively stable 4.7 cm long reference cavity, and studied the amplitude noise at the slope of a resonance. We observe noise levels which correspond to a FWHM laser linewidth of <300kHz within 1\,s. The observed frequency noise would lead to amplitude noise of $3\times 10^{-3}\,$rms when the microcavity transmission is observed at maximum slope of the resonance. This is more than one order of magnitude smaller than the smallest amplitude noise observed when probing the stabilized cavity. It is thus a small but relevant contribution to the observed noise, which limits the observed stability in particular at low frequencies.

The electronic lock achieves a length stability of $153\,$fm rms, is long-term stable, improves the mechanical stability by a factor $\sim 10^6$ compared to the free-running cavity, and achieves a noise level that is sufficient also for demanding applications.\\


\section{Photothermal bistability}

Due to the large finesse and the small waist of the optical mode ($w_0\simeq 2\,\mu$m), high intracavity intensities are easily reached, and light induced effects become prominent already for low incoupling powers $P_i$. E.g.\ for $P_i=1\,$mW and assuming unity fiber coupling and mode matching efficiency, the intracavity power amounts to $P_c = P_i T (F/\pi)^2 = 36\,$W and the peak intensity in an antinode reaches $I_\mathrm{max}=8P_c/(\pi w_0^2) = 2\,$GW/cm$^2$, on the order of damage thresholds for typical low-loss coatings. Here, $T = 1.2\times 10^{-5}$ is the mirror transmission for our coating. 

\begin{figure*}
\centering\includegraphics[width=0.9\textwidth]{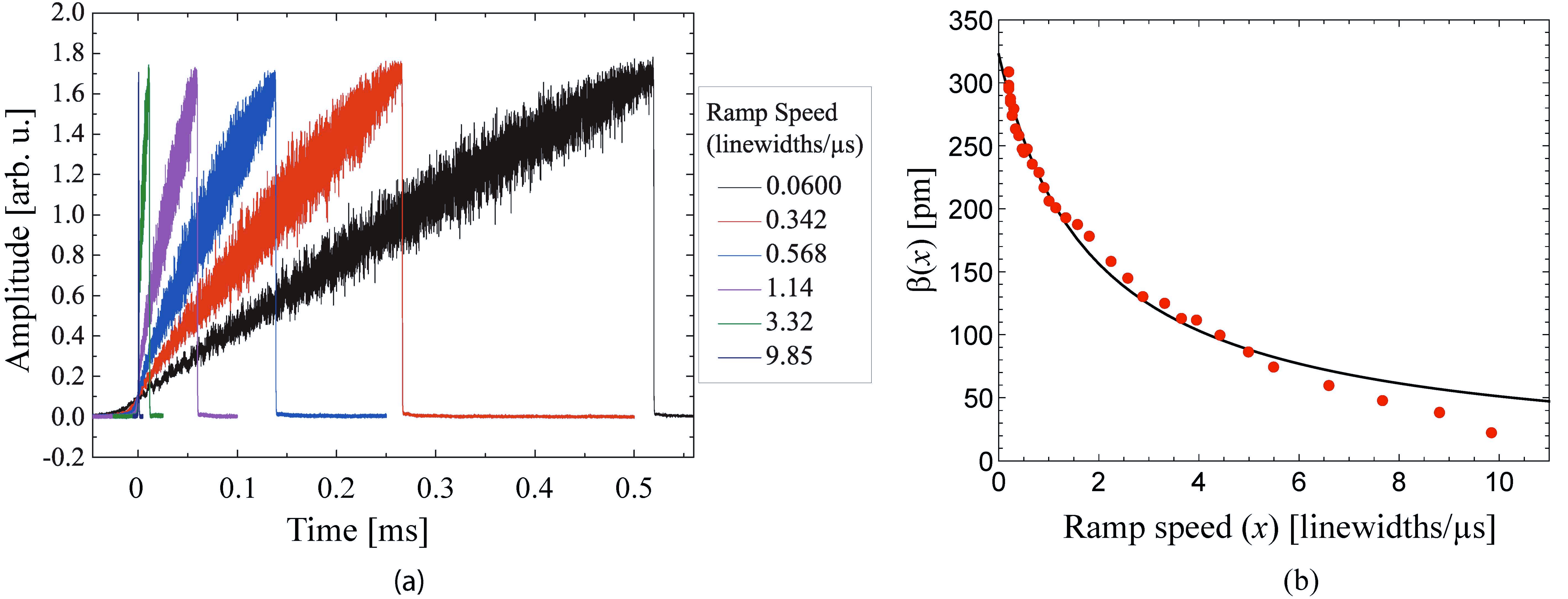}
\caption{(a) Examples of recorded lineshapes when ramping the mirror separation at different speeds across the resonance for an intracavity power of 38\,W. (b) Evaluated line shift $\beta$ as a function of ramp speed. The solid line is a fit of Eq.\ \ref{eq:fit} with $\beta_{\mathrm{ad}}=0.30$\,nm and $\tau = 0.5$\,$\mu$s.
}
\label{fig:bistability}
\end{figure*}

We observe photothermal bistability \cite{An97,Braginsky99,Rokhsari04,Hunger10b,Farsi12} due to mirror absorption for intracavity powers $P_c > 1.5$\,W. Figure \ref{fig:bistability}(a) shows ramp-speed-dependent lineshapes, where we scan the mirror separation to increasing length and probe the resonance with a fixed laser frequency. It can be seen that thermal expansion leads to self-stabilization on the red side of the resonance (i.e.\ where the cavity resonance frequency is higher than the probe laser frequency). Here, thermal expansion of the coating compensates mechanical length changes and thereby keeps the cavity close to resonance. This effect has previously been studied in the context of stabilization of toroidal resonators \cite{Carmon04,McRae09} and monolithic fiber-based Fabry-Perot microcavities \cite{Gallego16}. Scanning the cavity in the opposite direction causes a narrowing of the observed resonance shape, as heat induced expansion increases the scanning speed.

Absorption of radiation is significant only in the outer $4 - 6$ layers of the dielectric mirror coatings where the optical power is high. The heating rate of these layers follows changes in the intracavity power instantaneously, while heat conduction to lower lying coating layers and the substrate requires finite time. When modulating the intracavity power, the amount of material that contributes to the resulting temperature modulation depends on the modulation frequency $f$, leading to a dependence of the thermal expansion amplitude $\propto 1/f$ at high frequency \cite{Cerdonio01,Black04}.

We investigate the achievable bandwidth and amplitude of thermal self-stabilization for our cavity. Therefore, we measure the resonance lineshape as a function of the ramp speed $x$, see Fig.\,\ref{fig:bistability}(a), and evaluate the respective shift of the line peak $\beta(x)$, see Fig.\,\ref{fig:bistability}(b). We estimate the thermal reaction time constant $\tau$ and the adiabatic thermal lineshift $\beta_{\mathrm{ad}}$ from a simplified version of the model described in \cite{An97} and fit the data with a hyperbolic decay
\begin{equation}
\beta(x) = \frac{\beta_{\mathrm{ad}}}{1 + x \tau}.
\label{eq:fit}
\end{equation}

Within a limited parameter regime ($x \simeq 0\, ..\, 5$ linewidths$/\mu$s), good agreement between model and data is found. We extract the reaction time constant from the fit and find $\tau = 0.5\,\mu$s. The inverse, $f_\mathrm{th}=1/ (2\tau) = 1\,$MHz, gives a measure for the bandwidth that can be expected for photothermal resonator length stabilization. The value is comparable to the characteristic frequency of the coating \cite{Black04}, $f_c = 2\kappa_c/(\pi \rho_c c_c w_0^2 (\ln 2)^2) = 900\,$kHz when assuming bulk values for the thermal properties of the coating (average density $\rho_c = 3215\,$g/cm$^3$, average thermal conductivity $\kappa_c=5\,$W/(m\,K), average specific heat $c_c=680\,$J/(kg\,K) for Ta$_2$O$_5$ and SiO$_2$). The characteristic frequency separates between the weakly frequency dependent low-frequency behavior and the $1/f$ high-frequency roll off of the thermal expansion. From the coating thickness ($t_c = 6.6\,\mu$m) we estimate that for frequencies larger than $f_0=\kappa_c/(\pi\rho_cc_ct_c^2) \approx 20\,$kHz, the coating dominates the thermal expansion, while the substrate does for lower values \cite{Black04}.

The constant $\beta_{\mathrm{ad}}$ denotes the lineshift for adiabatic scanning conditions ($x \rightarrow 0$) and thus the maximum amplitude of cavity length changes due to thermal expansion. From the fit, we obtain $\beta_{\mathrm{ad}} = 0.30\,$nm. Alternatively, its value can be determined from material constants and geometry parameters of the resonator \cite{An97} according to $\beta_{\mathrm{ad}} = 4 \alpha_s C_0 A P_c /(\pi\kappa_s)$. With $C_0 =  3.1$ a constant depending on the mode and mirror geometry, $A = 6 \times 10^{-6} $ the mirror absorption, $P_c = 38\,$W as determined from the resonant cavity transmission $P_t = T P_c$, $\alpha_s = 5 \times 10^{-7}\,\rm{K}^{-1}$ the thermal expansion coefficient and $ \kappa_s = 1.4\, \rm{W}/\rm{m\,K}$ the thermal conductivity of the SiO$_2$ substrate respectively, we calculate a value $\beta_{\mathrm{ad}} = 0.32\,$nm, in good agreement with the experimental observation.
We note that a more detailed picture of the frequency dependent photothermal response can be obtained by modulating the absorbed laser power \cite{DeRosa02,Black04} and applying the model developed in \cite{Cerdonio01}, which will be subject of future work.\\

\begin{figure}
\centering\includegraphics[width=0.6\textwidth]{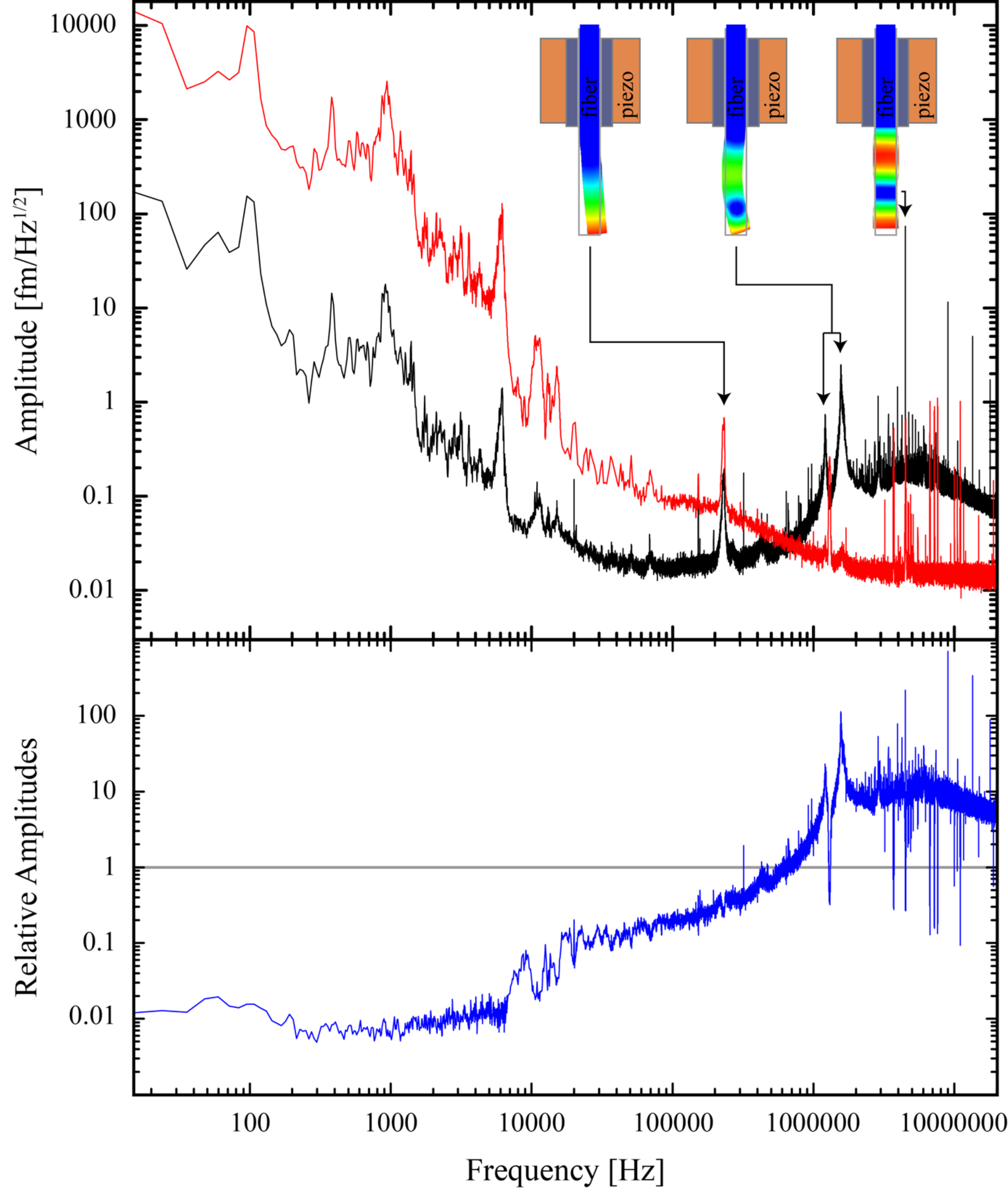}
\caption{Upper graph: improvement of a low gain electronic stabilization (red) by photothermal self-stabilization (black). The rms noise is 400\,pm (6.49\,pm) for the red (black) spectrum. Mechanical resonances of the optical fiber end are assigned to spectral features. Lower graph: Ratio of noise amplitudes $a_\mathrm{pt}(f)/a_0(f)$. Low frequency mechanical noise is suppressed by up to a factor 100, while high frequency oscillations are amplified at high intracavity power.}
\label{fig:thermal_lock}
\end{figure}


\section{Photothermal resonator length stabilization}

We now discuss photothermal self-stabilization of the microresonator. To increase self-stabilization activity at low frequencies, the gain of the electronic feedback is reduced. As a reference, we record the amplitude spectral density $a_0(f)$ for low intra-cavity optical power of a few ten milliwatts, see the red trace in Fig.\,\ref{fig:thermal_lock}. It is acquired with the 856\,nm laser to which the resonator is stabilized (and not the 780\,nm probe laser), because the fluctuations of the mirror separation are larger than the resonance width $\delta z$ of the 780\,nm laser. We then switch on the 780\,nm laser and tune it to the self-stabilizing slope of the resonance and launch an intra-cavity power of approximately 34\,W. The resulting spectrum $a_\mathrm{pt}(f)$ is shown in black. We observe a decrease in the noise spectral density by approximately two orders of magnitude for low frequencies, and a reduction of the integrated (10 Hz -- 20 MHz) rms noise by a factor 66. 

The ratio of the noise amplitudes of both spectra $a_\mathrm{pt}(f)/a_0(f)$ is shown in the lower part of Fig.\,\ref{fig:thermal_lock}. 
Photothermal response reduces the amplitiude of mechanical vibrations with a spectrally flat gain factor $a_\mathrm{pt}(f)/a_0(f)\sim 0.01$ for frequencies up to 10 kHz and shows a roll-off for higher frequency. The frequency range of photothermal self-stabilization extends up to 700\,kHz, more than two orders of magnitude larger than the bandwidth of the electronic feedback. Above 700\,kHz, high frequency mechanical noise and narrow resonances are excited, effectively limiting the performance of the photothermal lock.

Modal FEM simulations of the mechanical modes of the fiber end, with a measured protrusion of 0.63\,mm, predict mechanical resonances which fit very well to the narrow features seen in the spectra. Insets in Fig.\,\ref{fig:thermal_lock} assign the simulated modes to the main spectral features at high frequencies. While the fundamental bending mode of the fiber (left inset) at 250\,kHz experiences damping, the other resonances are amplified for large intracavity power. The center inset shows the first higher bending mode (n=2), which appears as a doublet (in-/out of plane) in the spectrum at 1.3 and 1.7 MHz, most probably due to the non-symmetric mounting of the fiber. The right inset shows a compression mode matching the resonance at 4.5\,MHz, where the center of mass oscillates along the optical axis and the diameter breathes. Sharp resonances at 9, 13.5, and 18\,MHz exactly match harmonics of the compression mode.

We attribute the strong excitation of the 4.5\,MHz mode to parametric amplification caused by time-lagged photothermal response, and the broadband high-frequency mechanical gain to radiation pressure excitation.
To understand the various contributions, we discern between a low-frequency in-phase component and a phase-lagged component at the frequency of mechanical resonances for the photothermal response \cite{Metzger08}. In addition, radiation pressure is expected to contribute to the observed mechanical response. The low-frequency influence of radiation pressure and photothermal expansion are opposite, where the former pushes the mirros apart and the latter reduces the mirror separation in the presence of an intracavity field. Consequently, for a red detuning of the laser with respect to the cavity, photothermal stabilization occurs, while radiation pressure leads to amplification, causing mechanical gain in the measurement shown in Fig.\,\ref{fig:thermal_lock}. The bandwidth of the radiation pressure effect is given by the cavity linewidth (115\,MHz) and is thus significantly larger than that of the photothermal response ($f_{th} = 1\,$MHz). Radiation pressure can thus become dominant at frequencies above the thermal response bandwidth, and can explain the increased noise level observed in Fig.\,\ref{fig:thermal_lock}. We note that the in-phase radiation pressure effect is typically neglected in cavity-optomechanical studies.
The opposite response is expected for mechanical resonances of the cavity when additional phase lag is present due to the finite bandwidth of the response. While radiation pressure provides damping in this regime, photothermal expansion leads to parametric excitation of mechanical motion. For the resonances observed in our measurement, the phase lag is negligible for radiation pressure, and little damping is expected. In contrast, photothermal expansion acquires a significant phase shift above $f_{th}$, and mechanical resonances with larger frequency are parametrically driven. This effect is expected to be the origin of the excitation of the resonance at 4.5\,MHz.\\


\section{Photothermal parametric instability}

\begin{figure*}
\centering\includegraphics[width=0.9\textwidth]{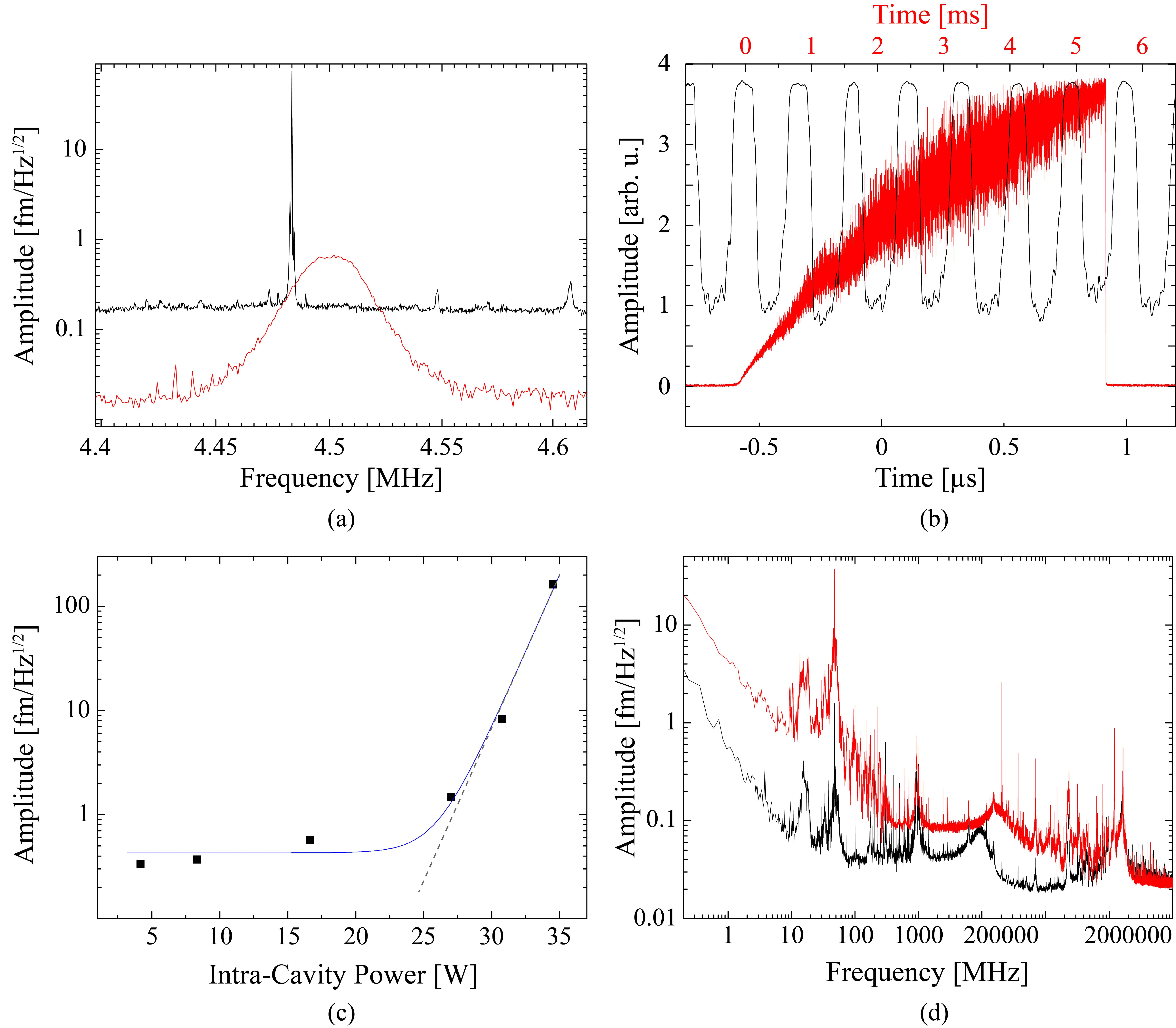}
\caption{(a) Compression mode at 4.5\,MHz for intracavity powers of 30.7\,W (black) and 60\,mW (red). (b) Transmission time traces of the stabilized resonator (black) and when increasing the cavity length across a resonance (red). Horizontal axis are color coded. Peak intracavity power is 38\,W. (c) Power dependence of the rms amplitude of the compression mode (black data points).The solid line is a fit. (d) Noise spectral densities with (black) and without (red) photothermal self-stabilization in an optimized setting. An intracavity power of 3.7\,W reduces the rms amplitude from 153\,fm to 114\,fm rms (integrated from 0.1\,Hz to 10\,MHz).}
\label{fig:4MHz_details}
\end{figure*}

In the following, we characterize the amplification of the compression mode in more detail to evaluate the compromise between photothermal self-stabilization and parametric instability.
Figure \ref{fig:4MHz_details}(a) shows the resonance with higher resolution for negligible and large intracavity power, corresponding to thermal and parametrically driven mechanical motion respectively. Parametric excitation increases the amplitude by more than two orders of magnitude and the linewidth narrows from its intrinsic width of $30\,$kHz to below $30\,$Hz, limited by the resolution of the measurement.

Figure \ref{fig:4MHz_details}(b) shows the bistable lineshape as previously shown in Fig.\,\ref{fig:bistability} together with a time trace of the stabilized resonator transmission. At a peak intracavity power of 38\,W, oscillations are driven over almost 3/4 of the resonance height, corresponding to an amplitude of approximately 1.8\,pm (peak-to-peak). One can see the non-sinusoidal shape, which manifests as harmonics in the frequency spectrum. Similar oscillations are observed for all time traces when the cavity length is increased across a resonance and high intracavity power is present, see e.g.\, increased noise in Fig.\,\ref{fig:bistability}.

We measure the amplitude of the compression mode for varying intracavity power, see Fig.\,\ref{fig:4MHz_details}(c). Therefore we record noise spectra at different intracavity powers and integrate the noise over the resonance to obtain the rms amplitude. Enabled by the high cavity finesse, we achieve a noise floor of $2\times 10^{-17}\,$m/$\sqrt{\mathrm{Hz}}$ around the resonance, allowing to easily resolve also the Brownian motion amplitude.
Parametric amplification is characterized by a threshold power, above which an exponentially growing oscillation amplitude with increasing intracavity power is expected. We extract the threshold power $P_0$ by fitting the data with a simple exponential function $ a(P) = a_{\rm{th}}(1 + e^{g (P-P_0)})$, from which we obtain $P_0=26.5\,$W, the proportionality constant  $g=0.75\,$W$^{-1}$, and a thermal amplitude of $a_{\rm{th}}=0.43\,$fm. The thermal amplitude is in agreement with an estimate via the equipartition theorem, $a_{\rm{th}} = \sqrt{k_B T/(m_{\rm{eff}}\omega_{\rm{m}}^2)}$, with $m_{\rm{eff}}=28\,\mu$g the effective mass of the fiber end taking part in the thermal motion and $\omega_m = 2\pi\times4.5\times 10^6\,$Hz.

We now reconsider the potential of photothermal self-stabilization with the presence of parametric instability. Experimentally, we set the intracavity power to a value that optimizes the combination of self-stabilization and parametric oscillation to obtain optimal cavity stability.
In Fig.\,\ref{fig:4MHz_details}(d), noise spectral densities of the resonator length with and without self-stabilization are compared. It can be seen that within the bandwidth of the electronic stabilization of 1.1\,kHz, the rms noise amplitude is reduced by approximately one order of magnitude, and the increased bandwidth of the photothermal effect improves the lock up to frequencies around 500\,kHz. Overall, the rms noise of the locked cavity is reduced by $34\%$, without detrimental parametric oscillations.

Low frequency noise dominates the spectra shown in Fig.\,\ref{fig:4MHz_details}(d), which we attribute to relative frequency drift between the free-running lock and probe lasers as well as the fluctuations of the parasitic Fabry-Perot of the cavity fiber. Furthermore, the radiation pressure force imprints the power and frequency noise of the probe laser onto the resonator length. A power and frequency stabilization of both lasers should thus reduce low frequency noise levels further.\\

\section{Conclusion}

We have demonstrated a robust stabilization scheme for fully tunable optical microcavities under harsh ambient conditions and achieved a stability of 114\,fm rms. We have furthermore shown that photothermal self-stabilization can improve an electronic lock and increase the bandwidth significantly. For a cavity with an intrinsic stability better than $\beta(f,P_c)$ for a given intracavity power, photothermal self-stabilization can provide full stabilization \cite{Gallego16}. We expect that an optimized mechanical design can be found that minimizes the impact of the parametric instability, opening the way for entirely photothermally stabilized resonators that are still fully tunable. Despite the high intracavity intensity, we have not observed any mirror degradation. In fact, we have rather observed an improvement of the cavity finesse from initially 130000 to 172000. A possible explanation is that thermal annealing occurs during locking at high intracavity power.

The demonstrated stabilization scheme has direct applications for portable, narrow-band single photon sources, for cavity-enhanced generation of indistinguishable single photons under ambient conditions \cite{Grange15}, cavity-enhanced spin-photon interfaces, scanning cavity microscopy \cite{Mader15,Huemmer16}, and cavity optomechanics \cite{FlowersJacobs12,Stapfner13}.
Furthermore, parametrically driven mechanical modes of optical fibers enable high-frequency modulation of the cavity transmission with large contrast and low phase noise, and could serve as a fiber-based all-optical frequency reference \cite{Hossein06,Tallur11}.\\


\section*{Funding}
Excellence Cluster Nano Systems Initiative Munich (NIM);
European Union's Seventh Framework Programme (FP7), (618078), (WASPS);
T. W. H\"ansch acknowledges funding from the Max-Planck Foundation.\\

\section*{Acknowledgments}
We thank Jakob Reichel, J{\"u}rgen Volz, and Albert Schliesser for helpful discussions.



\end{document}